
%

\magnification=1200
\baselineskip 20pt
\hsize 15truecm
\tolerance=1200
\pageno=1
\parindent=0mm
\raggedbottom
\overfullrule0pt
\hfuzz=5pt
\parskip\medskipamount
\hoffset=1.5truecm
\pretolerance=1600
\tolerance=1200
\hbadness=5000
\def\w{$W(100)\;$ }				  
\def\c{$c(2\times 2)\;$ }			  
\def\h{$c(2\times 2)-H\;$ }			  
\def\tr{\rm Tr\,}
\null
\font\lmr = cmr10 scaled \magstep 0
\font\bmr = cmr10 scaled \magstep 0
 2
\font\lbf = cmbx10 scaled \magstep 0
\font\bbf = cmbx10 scaled \magstep 0
 2
 0
\font\bit = cmti10 scaled \magstep 0
 2
 1
 1

\bmr
{\bbf

\centerline { THEORY OF ADSORBATE INDUCED }
\centerline { SURFACE RECONSTRUCTION ON }
\centerline { W(100) }
\centerline { }

}
\medskip
\centerline { Kari Kankaala $^{*,\dagger}$,
              Tapio Ala - Nissila$^{\ddagger,\dagger,**}$, and
              See-Chen Ying$^\ddagger$}
\medskip
\centerline { $^\ddagger$ Physics Department, Box 1843, Brown University, }
\centerline { Providence, RI  02912, USA}
\medskip
\centerline { $^*$Center for Scientific Computing, P.O. Box 40,}
\centerline { SF-02101 Espoo, Finland}
\medskip
\centerline { $^\dagger$ Department of Electrical Engineering,
              Tampere University}
\centerline { of Technology, P.O. Box 527, SF-33101 Tampere, Finland}
\medskip
\centerline { $^{**}$Research Institute for Theoretical Physics,}
\centerline { University of Helsinki,}
\centerline { Siltavuorenpenger 20 C, SF-00170 Helsinki, Finland}
\bigskip
{\lmr \baselineskip 12 truept
 \centerline {\lbf Abstract}

We report results of a theoretical study on an adsorbate induced surface
reconstruction. Hydrogen adsorption on a $W(100)$ surface causes
a switching transition in the symmetry of the
displacements of the $W$ atoms within
the ordered $c(2 \times 2)$ phase.
This transition is modeled by an effective Hamiltonian, where the hydrogen
degrees of freedom are integrated out. Based
on extensive Monte Carlo renormalisation group calculations we
show that the switching transition is
of second order at high temperatures and of first
order at low temperatures. This behavior is qualitatively explained
in terms of an $XY$ model where there is an  interplay
between four and eight fold anisotropy fields.
We also compare the calculated phase
diagrams with a simple mean field theory.
}

\vfill \eject

{\bbf 1. Introduction}

Structural phase transitions have been observed on a number of
surfaces.
One of the most extensively studied systems undergoing a surface
reconstruction is the clean $W(100)$
surface (see {\bit e.g.} [1,2,3] and refences therein).
It has a very rich phase diagram with a reconstructive phase transition
from a  $(1 \times 1) $ phase
into a $c(2\times 2)$ phase [1,4,5,6].
The microscopic structure of the low temperature phase was
unraveled early by Debe and King [5] and it has since then
been accepted that
this phase involves alternate displacements of the $W$ atoms along the
$<11>$ direction in the surface plane. A number of first principles
total energy calculations [{\bit e.g.} 7,8,9]
have verified the inherent instability
of the ideal $(1 \times 1)$ phase towards this reconstruction, and
the Debe-King
structure is found to have the lowest energy. There was some
dispute on
the nature of the high temperature phase, and thus
also on the details of the structural phase transition between these
two phases. Earlier, there was both
experimental [1,10,11,12] and theoretical [13]  evidence
which was interpreted in favor of an ordered high temperature phase where
the tungsten atoms would vibrate about their ideal bulk positions.
The other possibility was that the tungsten atoms would be displaced
from their bulk positions also in the high temperature phase but the
displacements would point into random directions, thus forming
a disordered phase [7,14].
Recent experimental
[6,15,16] and theoretical [17,18,19,20] studies have
corroborated the disordered
nature of the high temperature phase, and
it is now generally accepted that the transition is
of order-disorder type instead of a displacive one.

While the clean surface of $W(100)$ has been studied extensively
both experimentally and theoretically,
the influence of adsorbates on the reconstruction has been studied
to a far less extent and in most cases a thorough understanding of these
phenomena is missing.
Among the adsorbates, the adsorption of hydrogen has been studied in most
detail [1,4,10,22,23,26,27,44].
It was found that small amounts of hydrogen
enhance the inherent instability of the clean surface, and
increase the critical temperature of the order-disorder transition
[21].
The structure with adsorbed hydrogen was initially believed to
be the same as  the clean surface reconstructed \c phase
or very similar to it [1,5,22,23].
It was later shown [10] and verified [4,44]
that the hydrogen adsorbed surface has a different structure when
the coverage exceeds $\theta \simeq 0.1$.
Namely, the clean \c has a $p2mg$ symmetry whereas
the hydrogen rich (coverage $\theta > 0.1$)
\h phase has a $c2mm$ symmetry.
The symmetry change was cleverly verified by Griffiths {\bit et al.}
[24] in a low energy electron diffraction (LEED)
spot intensity study.
They studied the difference in the relative intensities of
two half-order beams, namely $(\bar {1\over 2},{1\over 2})$ and
$(\bar {1\over 2},\bar {1\over 2})$ at a normal incidence.
The difference between these two beams is finite in the
reconstructed clean surface $p2mg$ phase, and
it vanishes when the entire surface has switched into the $c2mm$
phase. This symmetry switching was {\bit initiated} at
a coverage of $\theta \simeq 0.04$ and completed at $\theta \simeq
0.16$. Similar symmetry switching has been observed in an infrared
spectroscopy study [25]. The $c2mm$ phase is interpreted
as due to a structure in which the $W$ atoms have alternate displacements
along the $<10>$ or $<01>$ directions.

The hydrogen induced switching  of the $W(100)$
surface between these two structures is the main subject of the present
work. Our study is motivated by theoretical
[26] and experimental [24] indications that the
switching transition could be of first order at low temperatures, instead
of being an $XY$ like continuous transition [27].
We will first briefly discuss the theoretical model for this work.
For the clean surface,
we adopt a lattice dynamical Hamiltonian
which has recently been shown to describe the critical properties of
the $W(100)$ surface
very accurately [17,18,19,20]. It corresponds to
an $XY$ model with an intrinsic cubic anisotropy field.
To this model then, we add the short range interactions of hydrogen with
its nearest neighbor $W$ atoms.
The hydrogen degrees of freedom are then
explicitly integrated out leaving an effective Hamiltonian.
This new Hamiltonian is shown to contain anisotropy fields
of all orders allowed by symmetry, and we argue that
the  nature of the switching transition at low temperatures
is dictated by
the interplay between effective fourth and eighth order anisotropy fields.
We then proceed to calculate the details of the phase diagram, using
first simple mean field arguments to locate the line of switching
transitions on the $(T,\mu)$ - plane. To study the nature of the
transition in detail, we present results of
extensive Monte Carlo Renormalisation Group calculations.
These results show that at high temperatures the transition
is continuous, while below a multicritical point $T_t$ it becomes
discontinuous. Finally,
we summarize our results and discuss the relevance of our work to experimental
studies. Preliminary results have been previously reported [28].

{\bbf 2. Theoretical Models }

As discussed above, it has been experimentally verified that hydrogen induces
a {\bit switching} transition where the
$H$ atoms adsorbed on bridge sites pin the
displaced $W$ atoms towards themselves resulting in displacement orientations
along either $<10>$ or $<01>$ direction, instead of the clean surface
$<11>$ direction.
This reconstruction has been subject
to theoretical investigations ({\bit e.g.} [3,8,9,26,27,29,30,31])
but to a far less extent than the clean $W(100)$ surface.

The models discussed below, as well as our model
belong to the class of so
called lattice dynamical models which are written in terms of lattice
displacements.
When studying the driving forces for the transition, microscopic details
have to be included in the model. However, lattice dynamical models
are usually sufficient to study the critical properties of the
systems under consideration. The role of the detailed electronic degrees of
freedom in these models has been discussed by various authors
[3,32,33].

{\lbf 2.1 Lattice Dynamical Models}

To study the effect of an adsorbate on the switching transition
on the $W(100)$ surface,
Lau and Ying [27] suggested a lattice dynamical model Hamiltonian

$$ {\rm H} = {\rm H}^{\rm clean} + {\rm H}^{\rm ad}. \eqno (1)$$

In their work, the clean surface term ${\rm H}^{\rm clean}$ is
of the type suggested by Fasolino {\bit et al.}
[34], with the details left unspecified. The crucial point is that for
the clean $W(100)$ surface,
it has the symmetry of an $XY$ model with cubic anisotropy.
The intrinsic anisotropy of the clean surface is such that the $W$ atoms
are displaced along the $<11>$ direction in the reconstructed \c phase.
The adsorbate part $ {\rm H}^{\rm ad}$ in Eq. (1) was chosen to be

$$ {\rm H}^{\rm ad} = {1 \over 2} \sum_{i',j'} J_{i'j'}n_{i'}n_{j'} +
            \sum_{i,i'} n_{i'}v({\vec R}_{i}^0 + {\vec u}_i - {\vec R}_{i'})
            \eqno (2) $$

which describes the effects of hydrogen on the structural transition.
In Eq. (2),
$n_{i'}$ is an occupation number, and $J_{i'j'}$ is an interaction term
between hydrogen atoms. The second summation describes the interaction
between hydrogen and tungsten atoms as a sum of pair potentials $v$
where
 ${\vec R}_{i'}$ is the coordinate of a hydrogen adsorption site and
${\vec R}_{i}^0$'s denote the
ideal bulk positions of the surface tungsten atoms. The addition of
${\vec u}_i$ describes the spontaneous displacements of the surface atoms.

Lau and Ying [27] started from Eq. (2)
and found that the adsorbate induces a cubic anisotropy field that opposes
the intrinsic clean surface one. This adsorbate induced anisotropy will
then increase in magnitude
with coverage leading to the change of sign of the total
anisotropy field. Hence the  switching occurs
from the phase in which $W$ atoms are displaced along the  $<11>$
directions to a phase in which they are displaced along the  $<10>$
directions, favored by the opposing cubic term.
This conclusion was supported by electronic band structure calculations
[29] where it was suggested that even
at low hydrogen coverages it would be energetically favorable
for the surface atoms to align along the $<10>$ or $<01>$
directions in the low temparature \c phase with hydrogen atoms
on bridge sites. Fasolino {\bit et al.} [35] suggested
the switching to occur at a coverage of $\theta \simeq 0.044
\simeq ({a \over \xi})^2$
where {\bit a} is the lattice constant and $\xi $ is the
surface coherence length.  This value for switching agrees with
LEED studies [24].

Ying and Roelofs [3,36] developed the model further by choosing
for the clean surface Hamiltonian

$$ \eqalignno { {\rm H}^{\rm clean} =
     &\sum_i \Bigl \{ {A \over 2}u_i^2 + {B \over 4} u_i^4
      + 8 h^0_4u_{ix}^2u_{iy}^2 \cr
     &+ C_1 (u_{ix}u_{jx} + u_{ix}u_{kx} +
      u_{iy}u_{jy} + u_{iy}u_{ky}) \Bigr \}, &(3)\cr} $$

where $u_i$ is the displacement of the surface $W$ atom at the $i^{th}$
lattice site, and $u_{ix}$ and $u_{iy}$ are the $x$ and $y$ components of
the displacement, respectively. $A$ and $B$ are coefficients of a simple
double well potential, $C_1$ is the nearest neighbor interaction term,
and  $h^0_4$ is the clean surface cubic anisotropy field. For the clean
\w surface $h^0_4 < 0$ which aligns the displaced
$W$ atoms along the $<11>$ directions.
Summation $i$ in Eq. (3) goes over all lattice sites.

Using (2) and (3), Ying and Roelofs obtained an effective
${\rm H}^{\rm eff}$
to describe the hydrogen effects by integrating out the hydrogen degrees of
freedom in the mean field approximation,
and  working with a constant coverage description.
However, this description breaks down if the transition happens to be of
first order as we shall discuss below. Also, the approximation of retaining
only the lowest order anisotropy fields is not justified for strong
$H - W$ coupling as addressed below.
In a closely related computer
study,
Sugibayashi {\bit et al.} [26] used an extended anisotropic $XY$ model
to describe the clean surface, with a
two - body lattice gas model for the adsorbate part.
They presented results from their computer simulations for
the switching transition, but were not able to conclude the
nature of the transition at low temperatures. Their results
seem to indicate, however, that there is a phase coexistence region at
low temperatures between the $<11>$ and $<10>$  phases, which would
be strong evidence for a first order transition.

{\lbf 2.2 New Model for H/W(100)}

To be able to study a possible first order regime, and to give
a more accurate description of the low temperature phase
we have undertaken simulations of a new lattice dynamical model.
The clean surface contribution in our model is that
described by Eq. (3) to which we have added an additional
eight fold anisotropy field:

$$ \eqalignno { {\rm H}^{\rm clean} =
     &\sum_i \Bigl \{ {A \over 2}u_i^2 + {B \over 4} u_i^4
      + 8 h^0_4u_{ix}^2u_{iy}^2 \cr
     &+ C_1 (u_{ix}u_{jx} + u_{ix}u_{kx} +
      u_{iy}u_{jy} + u_{iy}u_{ky}) \cr
     &+h^0_8\cos(8\phi_i) \Bigr\}, &(4)\cr} $$

where $\phi_i = \arctan(u_{iy}/u_{ix})$
is the displacement angle. To calculate the total
surface - adsorbate partition function, we will make the simplifying
assumption of neglecting direct $H-H$ interactions. Thus, we can
write

$$ Z = \tr _{\{n_{i'}\},\{u_i\}}
       \exp [ -\beta ({\rm H}^{\rm clean} + {\rm H}^{\rm ad})]
   \eqno(5) $$

with ${\rm H}^{\rm ad} = \sum_{i'} n_{i'} (V_{i'}-\mu)$, where
the summation $i$ goes over all lattice sites and $i'$ over
all bridge sites. As usual $\beta = 1 /(k_B T)$,
$n_{i'}$ is the actual hydrogen
occupation number at a bridge site $i' $, $V_{i'}$ is the interaction
potential between hydrogen and surface tungsten atoms, $\mu$
is the chemical potential, and $u_i$ is the displacement vector of a
tungsten atom.
Next, we sum over all hydrogen degrees
of freedom to obtain

$$ Z = \tr _{\{u_i\}} \exp [ -\beta ( {\rm H}^{\rm eff})],
\eqno (6) $$

where ${\rm H}^{\rm eff}$ is an {\it effective} Hamiltonian,
which can be obtained
in a straightforward fashion as

$$ {\rm H}^{\rm eff} = {\rm H}^{\rm clean} - k_B T\sum_{i'} \log \{ 1 +
            \exp [-\beta (V_{i'} - \mu)] \}.
\eqno (7) $$

The summation $i'$ goes over all bridge sites. They can also be
labelled as $i' \equiv (i, \nu)$ in which there are two bridge sites $\nu$
associated with each lattice site $i$. The interaction $V_{i'}$ is obtained
by expanding a pairwise interaction potential to lowest order in
the $W$ atom displacements. This gives us
$V_{i,\nu } = \sum_j {\bf \alpha}_{i \nu j}
         \cdot {\vec u}_j $ where the components of ${\bf \alpha}$ are
         given by ${\bf \alpha}_{i 1 1} = (-\alpha,0)$, ${\bf
         \alpha}_{i 1 2} = (\alpha,0)$, ${\bf \alpha}_{i 2
         1} =  (0,-\alpha)$, and
         ${\bf \alpha}_{i 2 2} = (0,\alpha)$, and ${\vec u}_j
         =(u_{jx},u_{jy})^T$, and $\alpha$ is an effective
         interaction parameter between hydrogen and tungsten atoms.
The relative positions of $H$ and $W$ atoms as well as displacement
vectors are illustrated in Fig. 1.

The Hamiltonian (7) is written in terms of the chemical potential
whereas the experiments are carried out at a constant coverage.
The description of the adsorbate in terms of the chemical potential
rather than the coverage allows us to better investigate the possibility
of both a first and a second order switching transitions.
The coverage can be calculated and is given by the expression

$$ \theta = < \sum_{i, \nu} { 1 \over {\exp [\beta(V_{i, \nu} - \mu)] + 1} } >
   \eqno (8) $$

in which the summation ${i, \nu}$ goes over all bridge sites and the
average $<\ >$ denotes an average over configurations.
Since there are
two bridge sites for each $W$, the maximum allowed coverage is normalized
to the value of two.

{\lbf 2.3 Effect of the Anisotropy Fields in the Model for H/W(100)}

We can predict qualitatively the behavior described by ${\rm H}^{\rm eff}$
if we express it entirely in terms of the displacement angles
$\phi _i$ for fixed amplitudes
instead of the displacement vectors ${\vec u_i}$.
If we retain only the isotropic nearest neighbor coupling terms and
the leading anisotropy fields in the on - site potential we then
obtain a simplified Hamiltonian

$$ {\rm H}^{\rm eff}_{XY} = K \sum_{i,j} \cos (\phi_i - \phi_j) -
       h_4(T, \mu) \sum_{i} \cos 4\phi_i +
       h_8(T, \mu) \sum_{i} \cos 8\phi_i .   \eqno (9) $$

The anisotropy fields $h_4(T,\mu) = h^a_4(T,\mu) + h^0_4$
and $h_8(T,\mu) = h^a_8(T,\mu) + h^0_8$ where superscripts $0$ and
{\bit a} refer to clean surface and adsorbate induced terms,
respectively.
This Hamiltonian is an $XY$ model with both four and
eight fold anisotropies.

The form of ${\rm H}^{\rm eff}$ together with the dependence of $h_4$
and $h_8$ on $T$ and $\mu$ has important consequences for the expected
behavior of the $H/W(100)$ system since
renormalisation group analysis [37]
tells us that the four fold field is relevant at all temperatures
$T < T_C$. For small values of $\mu$, $h_4(T,\mu) < 0$
which favors the orientation of the $W$ atoms along
the $<11>$ directions. However, since  $h^a_4(T,\mu)$ is a monotonically
increasing function of $\mu$ (for a fixed $T$), there will be
a point $(\mu_{\rm sw},T_{\rm sw})$ where $h_4(\mu_{\rm sw},T_{\rm sw})=0$.
Beyond this, $h_4(T,\mu)>0$, and the minimum energy state is
now given by $W$ atoms displaced along the $<10>$ directions.
This simple analysis predicts that the
switching transition in the $(T,\mu)$ - plane is determined
by the condition $h_4(\mu_{\rm sw},T_{\rm sw})=0$, and it is an $XY$
transition of Kosterlitz - Thouless type [37,38,39].
Physically, for the clean surface, the intrinsic four fold field
$h_4^0 < 0$ indicating a preference of the displacements for the
$W$ atoms along the $<11>$ directions. As $H$ is adsorbed, it sits
on the bridge sites and the $H-W$ interaction energy is minimized if locally
the displacements of the nearest neighbor $W$ atoms switch and point
along the $<10>$ or $<01>$ directions. This local distortion has a finite
range. Globally, the average $h_4(T,\mu)$ is still negative for
low hydrogen coverages. However, the effective $h_4(T,\mu)$ now
increases with the coverage or the chemical potential, reflecting these
local distortions around each $H$ atom.

To understand the nature of the switching transition in more
detail, we must also
consider the effect of the eight fold anisotropy field
$h_8(T,\mu)$ which is induced by the $H-W$ interaction term in (7).
It is well known that the eight fold field is relevant at low temperatures
but above some $T_t < T_C$ it becomes irrelevant [37].
When the eight fold field is negative, it favors
orientations  $\phi = n {\pi \over 4},\ n=0,...,7$, while for
$h_8(T,\mu) > 0$ the preferred orientations are $\phi = {\pi \over 8}
+ n {\pi \over 4},\ n=0,...,7$. This is demonstrated in Fig. 2.
Consider first a fixed, negative $h_8$. The eight fold field favors the
same directions
which are also favored by both positive and negative
$h_4(T,\mu)$. Thus, when $h_4(T_{\rm sw},\mu_{\rm sw}) \rightarrow 0$
from the negative side a relevant $h_8(T,\mu)$ field
keeps the system at the orientation $\phi = {\pi \over 4 }$.
However, as soon as the four fold field is positive and finite,
it dictates again the orientation of the system which
suddenly switches from   $\phi = {\pi \over 4 }$
to  $\phi = 0 $.
Thus, the transition is expected to be of first order at low temperatures
when the eight fold field is relevant and negative.
However, the situation is changed above $T_t$ where $h_8$ becomes
irrelevant. Then, at $h_4(T_{\rm sw},\mu_{\rm sw}) = 0$ the system is a pure
$XY$ model. The phase boundary thus has a multicritical point $T_t$
where $T_t$ and $T_{\rm sw}$ coincide for some $\mu$.
This is precisely the behavior for our choice of
${\rm H}^{\rm clean}$ as in (4) with a zero intrinsic eight fold field.
The reason is that the hydrogen adsorbate induced four fold and eight fold
fields are of the same sign, and thus compatible with each other.

For the case of a positive $h_8(T,\mu)$
which competes with the four fold field, more complicated behavior
can result below $T_t$. We expect then the phase boundary separating
the $<11>$ and $<10>$ phases to open up into
two Ising - like transitions with an intermediate phase in between [40,41].
This is a subject of a separate study.

It should also
be noted here that this discussion of four and eight fold fields
is strictly speaking
only qualitative for $H/W(100)$. As we discuss in the Appendix,
higher order fields are comparable in magnitude with $h_4$ and $h_8$
for a realistic $H - W$ coupling. Therefore,
the following mean field and Monte Carlo
Renormalisation Group analyses are based on the full effective
Hamiltonian of Eq. (7).
In addition, although the switching transition occurs at
$h_4(T,\mu)=0$, this four-fold field should be
a renormalized effective field
in which the short wavelength details have been
averaged away. Thus, even if we are able to extract correctly the bare
$h_4$ field in the Hamiltonian, it still would not allow us to
locate the switching point exactly. This is one more reason
why the numerical work is absolutely necessary.
The discussion above, however, is valuable in that
it nicely brings about the physical ideas behind the
change in the order of the switching transition.

{\bbf 3. Mean Field Estimate and the Choice of
         Parameters for the Switching Transition }

Before setting out to an extensive numerical study of our effective
Hamiltonian, it is useful to study the
mean field solution for the switching transition. The mean field
approximation is exact at $T=0$ and thus at least at low temperatures
it should yield reasonable results. At higher temperatures, the
temperature fluctuations affect the transition,
making the switching occur at a lower coverage or chemical potential
than the mean field theory predicts.

The energy difference between the two phases
that we denote by $\phi = {\pi \over 4}$ and $\phi = 0$ ($N$ is the
number of tungsten atoms), is:

$$ \Delta E
   = \Bigl [ {\rm H}^{\rm eff}(\phi = {\pi \over 4}) -
     {\rm H}^{\rm eff}(\phi = 0) \Bigr ], \eqno(10) $$

where
$$ \eqalignno {{\rm H^{\rm eff}}(\phi = {\pi \over 4}) =
   & 2Nh^0_4 + Nh^0_8 - k_BT\sum_{\nu,i} \log \bigl [
   {1 + \exp (-\beta (V_{\nu,i}(\phi =
   {\pi \over 4})-\mu))}\bigr ] \cr
   &+ N({A \over 2} + {B\over 4} -2C_1), & (11)\cr} $$

and

$$ \eqalignno {{\rm H^{\rm eff}}(\phi = 0) =
   &Nh^0_8 - k_BT\sum_{\nu,i} \log \bigl [
   {1 + \exp (-\beta (V_{\nu,i}(\phi = 0 ) - \mu))}\bigr ] \cr
   &+ N({A \over 2} + {B\over 4} -2C_1). & (12)\cr} $$

This can be simplified by noting that there are only four different kinds of
energy changes related to the bridge sites (total of $2N$) that can take place:

$$\eqalignno { \Delta E=
                &2Nh^0_4 \cr
		&-{N \over 2} k_BT \log \Bigl [
                 {{1 + \exp (-\beta (-\sqrt 2 \alpha - \mu))}
                   \over
                {1 + \exp (\beta \mu)}} \Bigr ]\cr
		&-{N \over 2} k_BT \log \Bigl [
                 {{1 + \exp (-\beta (-\sqrt 2 \alpha - \mu))}
                   \over
                {1 + \exp (-\beta (2\alpha - \mu))}} \Bigr ]\cr
		&-{N \over 2} k_BT \log \Bigl [
                 {{1 + \exp (-\beta (\sqrt 2 \alpha - \mu))}
                   \over
                {1 + \exp (\beta \mu)}} \Bigr ]\cr
		&-{N \over 2} k_BT \log \Bigl [
                 {{1 + \exp (-\beta (\sqrt 2 \alpha - \mu))}
                   \over
                {1 + \exp (-\beta (-2\alpha - \mu))}} \Bigr ]&(13)\cr}$$

In order to study in more detail the low temperature behavior of
the system, we set $\Delta E = 0$ and seek solutions to it.
These solutions become exact at $T=0$ in the absence of fluctuations.

At $T=0$, we get $\Delta E / N = {1 \over 2}\mu_{\rm sw} + 2h^0_4 + \alpha = 0$
in the interval $-2\alpha <\mu_{\rm sw} < -\sqrt{2}\alpha $. The solution
for $\mu_{\rm sw} $ is then $\mu_{\rm sw} = 2(-2h^0_4 - \alpha)$ where
$\mu_{\rm sw}$ is the value for the chemical potential where the switching
between $<11>$ and $<10>$ phases occurs. Thus, by fixing the values
of $\alpha$ and $h^0_4$ we can compare the prediction of the mean field
theory with the
experimentally observed values of coverage for the transition.
{}From the results we can also deduce a constraint between
the four fold anisotropy field $h_4^0$ and the interaction parameter
$\alpha$ as $4\vert h_4^0 \vert < (2 - \sqrt2 ) \alpha $
which restricts the choice of parameters, as will be discussed below.

In our model, there is also a solution for a higher chemical potential
which corresponds to the unphysical situation of
switching back to the $<11>$ phase at a higher coverage.
This is due to our omission
of direct interactions between hydrogen adatoms on adjacent bridge sites. A
direct repulsion between the adatoms on these sites would eliminate this
unphysical transition.
Experimentally, many complicated structures [4,25] have been
observed at high hydrogen coverages and these are outside the present scope of
study.

We have worked with two different sets of parameters. The first set (set I)
is very close to that of Yoshimori and coworkers [26] and was chosen so
that direct comparasion could be made with their results. We are particularly
interested in answering the question whether the transition is of first order
at low temperatures.
Their results indicates  this possibility
but the conclusions were unclear. A notable point in this set of parameters
is that the clean surface four fold field $h^0_4$ is very weak compared to
other similar studies [17,18,19,20] or total energy
calculations [8,9].

The other set of parameters (set II) is more consistent with  experimental
work and total energy calculations. A Hamiltonian based on this set of
parameters has been successfully used to explain many clean surface
phenomena [17,18,19]. However, to prevent a switching into displacements
along directions  other than $<10>$,
we need to set the eight fold field $h^0_8$ of Eq. 4 to a negative value.

In what follows, we work with dimensionless displacements $\tilde u$  and
temperature
$\tilde T$ defined as $u = u_s\tilde u, T = T_s\tilde T$ where $T_s$ and
$u_s$ are the scale factors.
The numerical values for the first set of parameters are
$\tilde A = Au_s^2/T_s = -2, \tilde B = Bu^4_s/T_s = 8.8,$
$\tilde C_1 = C_1u_s^2/T_s = 1.5 $, $\tilde h^0_4 = -0.1$, $\tilde
\alpha = 4.5$,
and $\tilde h_8^0 =0$.
The numerical values for set II are
$\tilde A = Au_s^2/T_s = -10, \tilde B = Bu^4_s/T_s = 40,$ and
$\tilde C_1 = C_1u_s^2/T_s = 3.75 $,  $\tilde h^0_4 = -1.85$,
$\tilde \alpha = 17$,
and $\tilde h_8^0 =-0.6 $.
The displacement amplitude $\tilde u = \sqrt {(4C_1 -A)/(B+8h_4)}
\simeq 1 $ for  parameter set I and is set to the value 1 for the
second set.
The critical temperature
for the clean surface ($\mu = -\infty$) is about $\tilde T_C \simeq 1.3 $ for
the first parameter set and about $\tilde T_C \simeq 2.3$
for the other set. The experimentally observed
values for $T_C \simeq 230K$
and $u \simeq 0.2 \AA$ yield the scale factors $u_s \simeq 0.2$ and
$T_s \simeq 175 $
for the first set and $T_s \simeq 100$ for the second set.
For simplicity, from here on throughout the paper we
will always use the scaled values for the parameters, {\bit e.g.}
$T$ for $\tilde T$.

The choice of $\alpha=4.5$ in the first
set of parameters follows of Roelofs and Ying [3]. In their
approach, the effective Hamiltonian was obtained by keeping terms only
to fourth order in the displacements of the $W$ atoms.
The various parameters including $\alpha $ were then
determined by comparing theoretical results with the experimental
data. As we show in the Appendix, the higher order terms are quantitatively
important and cannot be neglected. Indeed, the estimate for the clean surface
anisotropy $h^0_4$ in ref. [3] is about an order of magnitude
smaller than the current first principles results.
Thus, the value of $\alpha=4.5$ can also no longer be
viewed a reliable estimate. In the second set of parameters, the anisotropy
field is much stronger and
close to the first principles result. As a consequence,
the constraint $(2-\sqrt{2})\alpha > 4\vert h^0_4 \vert $
implies that a much larger value of $\alpha $ is
 required. Our choice of $\alpha=17 $
yields a switching coverage in agreement with
the experimental observation. It is also consistent with the infrared
vibrational spectroscopy data when the new larger value of the
anisotropy field is taken into account.

Finally, we should note that a
smaller value of $\vert h^0_4 \vert $ makes the transition more $XY$ - like
and thus more difficult to analyze numerically whereas a larger value
would decrease the fluctuations and result in more Ising like
behavior in the vicinity of the transition
with smaller finite
size effects.
Larger absolute values of the four fold field $h^0_4$ also hold
the system preferrably in the $<11>$ direction and thus would move
the switching to occur at a higher chemical potential. However,
the parameters within the model cannot be changed arbitrarily as
discussed above, and the value of $h^0_4$ imposes restriction to the value
of the interaction parameter $\alpha$. Increasing $h^0_4$
means increasing $\alpha $, and this leads the switching to
occur approximately at a constant coverage over a large temperature
range. The qualitative behavior of these two parameter sets
is the same both in the
$(T,\mu)$  and in $(T, \theta)$ planes (latter not sketched here).

{\bbf 4. Numerical Results}

To study the nature of the switching transition for the full
Hamiltonian of Eq. (7), we have carried out extensive
Monte Carlo renormalization group (MCRG) simulations [42].
By studying the cumulants of the order parameter,
we have been able to determine the order of the transition
both at low and high temperatures.
These results are supported by our additional studies
of the order parameter, its distribution function, coverage, and the
observation of strong hysteresis at low temperatures, as we will
eludicate below.

In the simulations, we have used the standard
Metropolis updating scheme. They were carried out using
a nonconserved order parameter and Glauber dynamics.
The amplitudes of the tungsten atom displacements $u_i$
were held fixed as it has been shown
recently [17,18] that displacement
amplitude fluctuations are not important for the critical
behavior of this system.
We studied mainly two system sizes, namely $64\times 64$ and
$32 \times 32$. Computational time involved with larger systems
becomes rather formidable.
The averages were computed over 50 000 configurations for
parameter set I, and over 20 000 for set II.

In the MCRG studies, the lattice is divided into blocks of size $L$,
and we then study the cumulants of the moments of the order parameter
distribution function. For this, we used two order parameters, namely

$$ <\Phi_x > = {1 \over {L^2}} \sum_{i}(-1)^{(i_x+i_y)}u_{xi}
\eqno (14) $$

$$ <\Phi_y > = {1 \over {L^2}} \sum_{i}(-1)^{(i_x+i_y)}u_{yi},
\eqno (15) $$

where $u_{xi}$ and $u_{yi}$ are the $x$ and $y$ components of the
displacement vector at the $i^{th}$ lattice site, and
$i_x$ and $i_y$ are the $x$ and $y$ coordinates of the
$i^{\rm th}$ lattice site in a block of size $L$.
The coefficient $(-1)^{(i_x+i_y)}$ is a phase factor.
$<\ >$ denotes an average over configurations,
and the summation goes over all lattice sites.
In the $<11>$ phase both (14) and (15) have finite values whereas in the
transition to
$<10>$ ($<01>$) phase  $<u_x>$ ($<u_y>$) vanishes.
There is no preference for one phase over the other.
By studying the behavior of the cumulants of these average displacement
components, we can probe the order of the switching transition.

The block cumulants used are defined by

$$ U_{Li} = 1 - { < {\Phi^4_i>_L} \over {3 <\Phi^2_i>_L^2} }, \eqno (16) $$

$$ V_{Li} = 1 - { {< \Phi^4_i>_L} \over {2 <\Phi^2_i>_L^2} }
           + { {<\Phi^6_i>_L} \over {30 <\Phi^2_i>_L^3} }, \eqno (17) $$

where $i$ refers to $x$ and $y$.
The variation of these two cumulants as a function of the block
size $L$ depicts a flow diagram analogous to that of a renormalisation
group method. It can be shown [42] that these cumulants
approach zero above $T_C$ as the block size increases. Below $T_C$,
both these cumulants tend to nonzero values
$U_L \Rightarrow U^* = {2 \over 3}, V_L \Rightarrow V^* = {8 \over 15}$,
and at a second order transition they tend to nontrivial values.

We also monitored the combined order parameter of Eqs. (14) and (15):

$$ <\Phi > = {1 \over {L^2}} \sum_{i}(-1)^{(i_x+i_y)}u_{xi}u_{yi},
\eqno (18) $$

For the ordered
$<11>$ phase this order parameter has a finite value,
and in the $<10>$ phase it vanishes.

If the transition is of first order, one of the block cumulants
is expected to flow towards the ordered state value of $2\over 3$
at all chemical potentials. As the transition occurs suddenly
between two ordered phases, one of the order parameter components does not
vanish at the transition
( $<u_y>$ for the $<01>$ phase and $<u_x>$ for the $<10>$ phase)
but abruptly changes from one ordered state value to the other.
The other component vanishes suddenly at the transition
( $<u_x>$ for the $<01>$ phase and $<u_y>$ for the $<10>$ phase).

One should note that if the transition is very weakly of first order,
it is very difficult to distinguish between it and a second
order transition. This is true in our case in the
vicinity of the multicritical point where the order of the
transition changes.

In the following, we will present numerical data only for one
parameter set (set I, Figs. 3 and 4) but will show the phase digram
for both parameter sets in Fig. 5. The reason for this is that results
for both sets are qualitatively very similar, and the quantitative
differences can be seen in the respective phase diagrams.
For parameter set I, the switching occurs at approximately
$\theta \simeq 0.02$ which is lower than the experimentally
observed value. For set II, we reproduce the switching at
approximately $\theta \simeq 0.1$ which is in good agreement
with experiments [24].

{\lbf 4.1. High Temperature MCRG Results}

Fig. 3 (a) shows the behavior for parameter set I
at high temperatures ($T=1.4$), and
confirms our predictions for the second order transition.
The switching transition in this case is from the $<11>$ state into
the $<01>$ state, i.e. $<u_x>$  vanishes.
The cumulants $U_{Lx}$ of $u_x$ show behavior typical for a second
order transition:
in the $<11>$ phase ($\mu < -12$) the cumulants approach
the fixed point of $U_{Lx} \approx 0.67$ and in the $<01>$ phase
$(\mu > 11.5)$ the cumulants approach zero.
We estimate the transition to happen in the region where
chemical potential is $-12 < \mu < -11$. Inside this range of
chemical potential, the cumulants approach a nontrivial
fixed point ($\mu \simeq -11.5$).
The problematic behavior of the cumulants at
$\mu = -14$ is attributed to large fluctuations
at this temperature and the vicinity of two
phase transitions (order - disorder transition and switching).

When studying the cumulants of $<u_y>$, we see that in the ordered
phase they approach the value of $U_{Ly} \approx 0.67$, and then near
the transition they flow to a nontrivial fixed point
at $\mu \in [-12, -11.5] $.
At larger values of  $\mu $ the cumulants again
approach the ordered state value. The errors have been
taken from the results from the largest lattice sizes.

Based on these results, we conclude that the switching transition at
$T_{\rm sw}=1.4$ occurs
at $\mu_{\rm sw} \in [-12, -11.5]$, and it is of second order.
Data for the parameter set II are very similar, and
result in a switching transition
for
$T=2.4$ ($< T_C$, high temperature) at $\mu_{\rm sw} \in [-28.4, -28.0]$.

{\lbf 4.2 Low Temperature MCRG Results}

In Fig. 3 (b) we depict the behavior of the cumulants
at $T_{\rm sw}=0.5$ again for the parameter set I.
At chemical potential values $\mu < -8.8$, the
cumulants of $U_{Lx}$ all approach the ordered phase value.
At  $\mu > -8.8$, the cumulants vanish. The cumulant behavior
of $U_{Ly}$ shows no change in behavior when passing through the
transition
but the cumulants approach the ordered state value of $0.67$
at all chemical potentials.
We interpret this as evidence of a first order phase transition
into the $<10>$ phase at this temperature. The strange behavior of $U_{Ly}$ at
$\mu = -8.8 $ is probably to be due to
strong metastable effects combined
with slow dynamics of the system at these low temperatures.

Based on similar arguments, the switching for set II occurs for $T=0.6$
at  $\mu_{\rm sw} \in [-26.7, -26.4] $. In this case the metastability
effects seem to be less severe, and the accuracy of simulations is
thus better even at relatively low temperatures.

\vfill \eject

{\lbf 4.3 Results from Other Quantities}

Futher qualitative support for the MCRG results can be found in our
coverage studies. In Fig. 4 the coverage as in Eq. (8)
is plotted as a function of temperature
for different values of the chemical potential.
If the transition were of first order a discontinuity
in the real coverage would be observed at the transition.
However, as we work with a finite system,
we observe only a very sharp rise in coverage
at the transition.

In the case of a second order transition, we would expect the
coverage to increase gradually when the transition region is passed.
The data in Fig. 4 is in qualitative agreement with the
prediction of a first order transition at low temperatures
and a second order transition a higher temperatures.
The abrupt increase in coverage is very pronounced at $T=0.2$.
Similar behavior was also observed for the order parameter.
The jump in coverage was very clear in simulations with
set II parameters in the first order regime, and more gradual at
higher temperatures reconfirming the qualitative similarity in
simulations for both parameter sets.

In addition, we have observed strong hysteresis and phase coexistence
(as determined from the order parameter distribution function) at
low temperatures to further support the scenario for a first
order transition. Hysteresis studies have not been performed
for the second parameter set.

{\lbf 4.4 The Phase Diagram}

Based on the mean field theory and our numerical
results, we sketch phase diagrams for both parameter
sets in Figs. 5 (a) and 5 (b)
for the line of switching transitions as
induced by hydrogen adsorption on
the $W(100)$ surface.
The temperatures for the $(1\times 1) \rightarrow c(2\times2)$
transition both for the clean surface ($\mu = -\infty $)
and for the low hydrogen coverages have
been determined from MCRG studies.
The solid order - disorder line is only a guide to the eye.

In both Figs. 5 (a) and 5 (b),
the dashed line denotes the mean field solution.
The solid line is the part of the switching transition where
it is expected to be of second order, and the dotted line depicts
the first order switching transition. The
agreement with the mean field line is very good
at low temperatures where the role of fluctuations
is not important. At higher temperatures the deviations are due to
temperature fluctuations which cause
the system to undergo a phase transition at lower chemical
potentials than our mean field results would imply.
The mean field result for the parameter set II at higher temperatures
deviates particularly strongly from the MCRG data.

The whereabouts of the multicritical point $T_t$
where the transition changes its order could not
be accurately pinpointed by our simulations. The locations of
the multicritical points depicted in Fig. 5 are only schematic.
We expect it to be somewhere between $0.5 < T_t < 1.0 $
for set I, and  $1.0 < T_t < 1.5 $ for set II.
The error bars shown depict the variance of numerical
results from a series of simulation runs.

{\bbf 5. Summary and Discussion}

To summarize, we have developed a model Hamiltonian to describe the
adsorbate induced effects on the $W(100)$ surface. The clean surface part
consists of a lattice dynamical Hamiltonian, which describes the
$(1 \times 1) \rightarrow c(2 \times 2)$ transition as studied by
Han and Ying [17,18]. To include
the hydrogen induced effects, we have considered
a simple model of interactions between $W$ and $H$ atoms, and
integrated out the hydrogen degrees
of freedom to obtain an effective Hamiltonian. This leads to an
$XY$ model with anisotropy fields of all orders allowed by symmetry.
In particular, the interplay between the fourth and eighth order anisotropy
fields indicates that the switching transition should be first order
at low
temperatures and of second order at high temperatures. This
prediction was confirmed by Monte Carlo
renormalisation group calculations, and further
corroborated by studies of other quantities.
We also used the MCRG studies together with a simple mean field theory
to map out the line of switching transitions on the $(T,\mu)$ plane.

We have carried out simulations with two rather different
sets of parameters, and
obtained qualitatively similar behavior for both sets. With set I,
we  have been able to reproduce results similar to those of
Sugibayashi {\bit et al.}, and re - interpret their findings in terms of
first and second order transitions. The transition coverage is
somewhat lower than experimantally observed. The second parameter
set is more in  agreement with total energy calculations,
and the results are qualitatively same as for the first set.
These both parameter sets reveal qualitatively similar behavior.
Thus, regardless of the strength of the four fold anisotropy field
in an experimental sample, a change in the order of the switching
transition should be observed as a function of temperature.

Experimental evidence from an infrared spectroscopy (IR) study
supports the scenario for a second order transition
at room temperature. In their study, Arrecis {\bit et al.} [25] observe
only one peak in their IR spectra over the coverage range
$\theta \in [0.044, 0.22]$. If there were coexisting phases present,
this would show as additional peaks in the spectra
corresponding to the different symmetries of the phases.
The LEED study of Griffiths {\bit et al.} was performed
at lower temperatures ($T\simeq 200 \ K$) although the
temperature was not held constant [24]. Their data shows indications
of a coexistence region between $\theta \in [0.05, 0.16]$
which could be interpreted in favor of a first order transition.
The temperature range in Ref. 24 is somewhat higher than our
predictions but is in qualitative agreement with our work.
Very recently, Okwamoto [43] has presented results of mean
field calculations
on the model of Lau and Ying [27] showing
the switching transition to be of first order at all temperatures below
the order - disorder line.
However, it should be noted that as the cubic anisotropy field vanishes
at the transition, Okwamoto's method is equivalent
to treating the pure $XY$ model with a mean field theory.
In this case it is well known that the mean field
treatment does not predict the order of the transition correctly as the
angular spin fluctuations destroy conventional long range order.

The major drawbacks in this study were the slow dynamics of the
model and the finite size effects. These  together made it impossible
for us within given computer time to locate the multicritical point
more accurately.
In addition, the adsorbate
induced part generates all allowed anisotropy fields which complicates
the analysis. From an experimental point of view, however,
the location of the multicritical point may vary from sample to sample,
depending on the possible intrinsic eight fold field. Thus
a more accurate theoretical determination of
$T_t$ may not prove to be necessary after all.

{\bbf Acknowledgements:}

Thanks to Mike Kosterlitz for many useful discussions, and to
Wone Keun Han for useful advice in simulations.
S.C.Y. and T. A-N. were supported in part by an ONR contract. K.K.
was supported in part by a contract between the Center for Scientific
Computing and the Academy of Finland, and grants from the Finnish
Academy of Sciences and the Finnish Cultural Foundation.
K.K. also acknowledges the hospitality of Brown University
where part of this work was carried out.
Authors acknowledge the computational resources provided by the
Center for Scientific Computing, Finland, Tampere University of
Technology, Finland, and Brown University, U.S.A.

\vfill \eject

{\bbf Appendix}

In this Appendix, we will discuss the hydrogen induced anisotropy fields
in more detail. We shall first write a series expansion for the part of the
Hamiltonian which is due to the adsorbate. We shall then expand the $H-W$
interaction potential in terms of the anisotropy fields, and discuss
the convergence of this expansion.

We start from the adsorbate induced part of Eq. (7)
and denote fugacity by $\gamma = \exp (\beta \mu )$.
The summation $i $ goes over
all lattice sites and $\nu $ over both bridge sites on each lattice site.
Notation is in all cases equivalent to that in the text.

$$ \eqalignno {{\rm H ^{\rm eff} - \rm H ^{\rm clean}}
   &= -k_BT \sum_{ {i},\nu }
      \log (1 + e^{ -\beta (V_{ {i},\nu } - \mu ) } ) &(A.1)\cr
   &= -k_BT \sum_{ {i},\nu }
      \Bigl [\gamma \exp (-\beta V_{ {i},\nu })
       - {{\gamma ^2} \over 2}\exp (-2\beta V_{ {i},\nu })
          \pm      ...\Bigr ] &(A.2)\cr
   &= -k_BT \sum_{i , \nu}
      \Bigl [\gamma \sum_n(-\beta)^n{{(V_{i, \nu})^n} \over {n!}}
      - {{\gamma ^2} \over 2}
        \sum_n(-2\beta)^n{{(V_{i, \nu})^n} \over {n!}}
       \pm ...\Bigr ] &(A.3)\cr
   &= -\sum_{ {i},\nu ,n} (-\beta)^{n-1}{{(V_{i, \nu})^n} \over {n!}}
      \Bigl [\gamma - {{2^n\gamma ^2} \over 2 }
      + {{3^n\gamma ^3} \over 3 } \mp ...\Bigr ] &(A.4)\cr
   &= -\sum_{ {i},\nu ,n} (-\beta)^{n-1}{{(V_{i, \nu})^n} \over {n!}}
      \sum_{k=1}^{\infty } \gamma ^k (-1)^{(k+1)}k^{(n-1)} &(A.5) \cr
   &\buildrel \rm def \over =
      -\sum_{ {i},\nu ,n} (-\beta)^{n-1}{{(V_{i, \nu})^n} \over {n!}}
      F_n(\gamma ,\mu) &(A.6) \cr } $$

Next, we study the effect of the $H-W$ interaction
terms $\sum_{i, \nu}(V_{i, \nu})^n$
which generate all the anisotropy fields induced by the
hydrogen degrees of freedom. Due to symmetry arguments, the odd powers
of ${\vec u}$ vanish. Let us now for the sake of argument
consider the term in Eq $(A.5)$ for $n=4$:

$$ \eqalignno { \sum_{i, \nu} (V_{i, \nu})^4 =
    \sum_{i ,\nu} \Bigl [
        &{\sum_j{(\bf \alpha}_{i \nu j}\cdot {\vec u}_j)}
         {\sum_k{(\bf \alpha}_{i \nu k}\cdot {\vec u}_k)}\cr
        &{\sum_m{(\bf \alpha}_{i \nu m}\cdot {\vec u}_m)}
         {\sum_n{(\bf \alpha}_{i \nu n}\cdot {\vec u}_n)} \Bigr ],
        &(A.7)\cr }$$

which reduces to

$$\sum_{i, \nu} (V_{i, \nu})^4
  = \sum_{i,\nu, j}\bigl [\alpha_{i \nu j}\cdot
    {\vec u}_j\bigr ]^4 \eqno (A.8) $$

when we consider only the on-site terms.

On the other hand, based on symmetry arguments we can readily
write the expansion for $(A.1)$ in a general form when we consider
contributions only from the on-site terms:

$${\rm H ^{\rm eff} - \rm H ^{\rm clean}}
   = \sum_i \sum_{k=0}^{\infty} h^a_{4k}(T,\mu) \cos (4k\phi_i)
     \eqno(A.9)$$

where $\phi_i $ is the displacement angle. The terms where $k=1$ and $2$
correspond to the analysis in Sec. 2.3.

When we now expand Eq. $(A.8)$
in terms of the displacement angle $\phi $, and note that
$\vec u = (u_x, u_y)^T, \vert \vec u \vert = u_0,$
$u_x = u_0\cos \phi $,  $u_y = u_0\sin \phi $ and $\vert
\alpha_{i \nu j} \vert = \alpha $, we get a contribution for
the four fold anisotropy field as

$$ -{{k_BT} \over {4!4}}
               \Bigl( {{\alpha u_0} \over {k_BT}} \Bigr )^4
       \Bigl [ \gamma - 8\gamma ^2 + 27\gamma ^3 - 64\gamma ^4
       + - ... \Bigr ] \eqno(A.10) $$

However, it is important to notice that this is {\bit not}
the total adsorbate induced field $h_4^a(T,\mu)$. Namely,
the evaluation of higher powers of $(V_{i,\nu})^n$ in Eq.
$(A.5)$ results in
contributions not only to $h_n^a$, but also to lower order
anisotropy fields.
These contributions
can be of the same order of magnitude as the leading terms which
severely hampers the analysis of these expansions. Thus, for
example, for the coefficient of the  four fold anisotropy field
we obtain

$$ \eqalignno {h_4^a(T,\mu) =
	& -{ {k_BT} \over {4!4}}
           \Bigl ({{\alpha u_0} \over {k_BT}} \Bigr )^4 F_4(\mu, T)
          - { {3k_BT} \over {6!8}}
           \Bigl ({{\alpha u_0} \over {k_BT}} \Bigr )^6 F_6(\mu, T) \cr
        & -{ {7k_BT} \over {8!16}}
           \Bigl ({{\alpha u_0} \over {k_BT}} \Bigr )^8 F_8(\mu, T)
          -{ {15k_BT} \over {10!32}}
           \Bigl ({{\alpha u_0} \over {k_BT}}
           \Bigr )^{10} F_{10}(\mu, T) \cr
        &  -{ {495k_BT} \over {12!1024}}
           \Bigl ({{\alpha u_0} \over {k_BT}} \Bigr )^{12} F_{12}(\mu, T)
            \pm ...  &(A.11) \cr } $$

where $F_n(\mu, T)$ is defined by $(A.5)$ and $(A.6)$.

To study analytically the interplay between the cubic and
eighth order anisotropy fields, and the vanishing of the total
$h_4(T,\mu)$ at the switching transition,
one must first ensure the overall convergence condition
$ \exp \{ -\beta (V_{ {i},\nu } - \mu ) \} \le 1$.
Unfortunately, for the values we have used for the parameters
$\alpha, \mu$ and $T$, this condition is not usually met
in the vicinity of the switching transition line.
We have also verified this numerically.
In addition, for the parameters that we use,
the term $\alpha u_0 / (k_BT)$ is never
less than one for physically reasonable temperatures, thus further
hampering the convergence of $(A.11)$.
The additional effect of the more complicated off-site terms,
which involve products of cosines at different lattice sites,
is also difficult to determine accurately but it seems evident
that they should be included in a quantitatively accurate calculation.

\vfill \eject

{\bbf References:}

\bigskip

\item{[1] }D.A. King,{ Physica Scripta} {\bbf T4}, 34 (1983).

\item{[2] }L.D. Roelofs and P.J. Estrup,
{ Surf. Sci.} {\bbf 125}, 51 (1983).

\item{[3] }L.D. Roelofs and S.-C. Ying,
{ Surf. Sci.} {\bbf 147}, 203 (1984).

\item{[4] }R.A. Barker and P.J. Estrup,
{ J. Chem. Phys.} {\bbf 74}, 1442 (1981).

\item{[5] }M.K. Debe and D.A. King,
{ J. Phys. C} {\bbf 10}, L303 (1977).

\item{[6] }I.K. Robinson, A.A. MacDowell, M.S. Altman, P.J. Estrup,
K. Evans-Lutterodt, J.D. Brock, and R.J. Birgeneau,
{ Phys. Rev. Lett.} {\bbf 62}, 1294 (1989).

\item{[7] }D. Singh, S.-H. Wei, and H. Krakauer,
{ Phys. Rev. Lett.} {\bbf 57}, 3292 (1986).

\item{[8] }D. Singh and H. Krakauer,
{ Phys. Rev. B} {\bbf 37}, 3999 (1988).

\item{[9] }C.L. Fu and A.J. Freeman,
{ Phys. Rev. B} {\bbf 37}, 2685 (1988).

\item{[10] }M.K. Debe and D.A. King,
{ Surf. Sci.} {\bbf 81}, 193 (1979).

\item{[11] }C. Guillot, C. Thuault, Y. Jugnet,D. Chauveau, R. Hoogewijs,
J. Lecante, Tran Minh Duc, G. Tr\`eglia, M.C. Desjonqu\`eres, and
D. Spanjaard,
{ J. Phys. C}{\bbf 15}, 4023 (1982).

\item {[12]} C. Guillot, M.C. Desjonqu\`eres,
D. Chauveau, G. Tr\`eglia,
J. Lecante, D. Spanjaard, Tran Minh Duc,
{J. Phys. C}{\bbf 15}, 4023 (1984).

\item{[13] }C.L. Fu, A.J. Freeman, E. Wimmer, and M. Weinert,
{ Phys. Rev. Lett.} {\bbf 54}, 2261 (1985).

\item{[14] } I. Stensgaard, L.C. Feldman, and P.J. Silverman,
{ Phys. Rev. Lett.} {\bbf 42}, 247 (1979).

\item{[15] } J. Jupille, K.G. Purcell, G. Derby, J.F. Wendelken, and D.A. King,
in {\bit Structure of Surfaces II}, edited by J.F. van der Veen
and M.A.Hove (Springer, Berlin, 1988), p. 463. and  J. Jupille,
K.G. Purcell, D.A. King, Phys. Rev. B {\bbf 39}, 6871 (1989).

\item{[16] }I. Stensgaard, K.G. Purcell, and D.A. King,
{ Phys. Rev. B} {\bbf 39}, 897 (1989).

\item{[17] }W.K. Han and S.-C. Ying,
{ Phys. Rev. B} {\bbf 41}, 9163 (1990).

\item{[18] }W.K. Han and S.-C. Ying,
{ Phys. Rev. B} {\bbf 41}, 4403 (1990).

\item{[19] }W.K. Han, Ph.D. Thesis, Brown University,
unpublished (1990).

\item{[20] }L. Roelofs, T. Ramseyer, L.L. Taylor, D. Singh, and
H. Krakauer, Phys. Rev. B, {\bbf 40}, 9147 (1989).

\item{[21] }R.A. Barker and P.J. Estrup,
{  Phys. Rev. Lett.} {\bbf 41}, 1307 (1978).

\item{[22] }T.E. Felter, R.A. Barker, and P.J. Estrup,
{ Phys. Rev. Lett.} {\bbf 38}, 1138 (1977).

\item{[23] }R.A. Barker, P.J. Estrup, F. Jona, and P.M. Marcus,
{ Solid State Comm.} {\bbf 25}, 375 (1978).

\item{[24] }K. Griffiths, D.A. King, and G. Thomas,
{ Vacuum} {\bbf 31}, 671 (1981).

\item{[25] }J.J. Arrecis, Y.J. Chabal, and S.B. Christman,
{ Phys. Rev. B} {\bbf 33}, 7906 (1986).

\item{[26] }T. Sugibayashi, M. Hara, and A. Yoshimori,
J. Vac. Sci. Technol. A {\bbf 5(4)}, 771 (1987).

\item{[27] }K.H. Lau and S.-C. Ying,
{ Phys. Rev. Lett.} {\bbf 44}, 1222 (1980).

\item{[28] }K. Kankaala, T. Ala-Nissila, and S.-C. Ying,
{ Phys. Scr.} {\bbf T33}, 166 (1990).

\item{[29] }D.W. Bullett and P.C. Stephenson,
{ Solid State Comm.} {\bbf 45}, 47 (1983).

\item{[30] }L. Lou, D.C. Langreth, and P. Norlander,
Surf. Sci. {\bbf 234}, 412 (1990).

\item{[31] }P. Norlander, S. Holloway, and J.K. N\o rskov,
Surf. Sci. {\bbf 136}, 59 (1984).

\item{[32] }W.L. McMillan,{ Phys. Rev. B} {\bbf 16}, 643 (1977).

\item{[33] }S.-C. Ying, In
{\bit Dynamical Phenomena at Surfaces, Interfaces
and Superlattices}, eds. Nizzoli, Rieder and Willis,
pp. 148--156 (Springer-Verlag, Berlin 1985).

\item{[34] }A. Fasolino, G. Santoro, and E. Tosatti,
{ Surf. Sci.} {\bbf 125}, 317 (1983).

\item{[35] }A. Fasolino, G. Santoro, and E. Tosatti,
{ Phys. Rev. Lett.} {\bbf 44}, 1684 (1980).

\item{[36] }S.-C. Ying and L.D. Roelofs,
{ Surf. Sci.} {\bbf 125}, 218 (1983).

\item{[37] }J.V. Jos\`e, L.P. Kadanoff, S. Kirkpatrick, and
D.R. Nelson,{ Phys. Rev. B} {\bbf 16}, 1217 (1977).

\item{[38] }J.M. Kosterlitz and D.J. Thouless,
{ J. Phys. C,} {\bbf 6}, 1181 (1973).

\item{[39] }J.M. Kosterlitz,{ J. Phys. C} {\bbf 7}, 1046 (1974).

\item{[40] }J.V. Selinger and D.R. Nelson,
{ Phys. Rev. Lett.} {\bbf 61}, 416 (1988).

\item{[41] }J.M. Kosterlitz, {\bit private communication} (1988).

\item{[42] }K. Binder,
{ Z. Phys. B} {\bbf 43}, 119 (1981).

\item{[43] }Y. Okwamoto, J. Phys. Soc. Japan {\bbf 60}, 2706 (1991).

\item{[44] }J.A. Walker, M.K. Debe, and D.A. King,
{  Surf. Sci.} {\bbf 104}, 405 (1981).

\vfill \eject

{\bbf Figure captions}

{\lbf Figure 1}

Here we show schematically the relative positions of $W$ and $H$ atoms on
the $W(100)$ surface. The solid dots ($\bullet$) denote the ideal $W$ atom
positions, and the vectors $\vec u_i$ denote the positions of the
displaced $W$ atoms. The hydrogen atoms reside on the bridge sites ($\times $)
$i' = (i, \nu)$ where $i$ denotes the tungsten atom lattice sites,
and $\nu = 1,2$ the two bridge sites related to each lattice site.
The $H$ atoms feel the $W$ atom potential of the nearest sites only,
{\bit e.g.} $V_{i1}=-\alpha u_{ix} + \alpha u_{kx}$, cf. Eq. (7). We have
also depicted the magnitudes of the displacements and the lattice
constant.

{\lbf Figure 2}

{\bit (a)}. The eight directions favored by the positive and
negative four fold anisotropy fields. The dotted directions are
favored by a negative $h_4$ and the solid lines show directions
favorable for a positive $h_4$. A negative eight fold field $h_8$ favors
all these directions.
{\bit (b)}. The eight directions favored by a positive eight fold
field. Compared to $(a)$ these directions are rotated by ${\pi \over 8}$.

{\lbf Figure 3}

{\bit (a)}. MCRG cumulant behavior typical for a second order
transition at $T=1.4$, and
{\bit (b)} for a first order behavior at low temperatures ($T=0.5$)
for parameter set I. See text for details.

{\lbf Figure 4}

Coverage (Eq. (8)) increases as a function of the chemical potential.
At elevated temperatures the increment takes place smoothly whereas at
lower temperatures an abrupt increase is seen. The latter would
indicate a possible first order transition at low temperatures as
discussed in the text. The calculation is performed for parameter set I.

{\lbf Figure 5}

The phase diagrams for $H/W(100)$ in the ($T,\mu$) plane,
based on numerical simulations of our model Hamiltonian
for parameter sets I {\bit (a)} and II {\bit (b)}
The solid dots denote the MCRG results.
The switching transition changes order at $T_t$ above which the
transition is of second order (solid line), and below which it is
assumed to be of first order (dashed line). The dotted line shows
the mean field result. The solid line between the ordered and disordered
phases in {\bit (a)} is only a guide to the eye.

\bye